\documentclass[12pt]{iopart}

\usepackage{graphicx}
\usepackage{dcolumn}
\usepackage{bm} 
\usepackage{amssymb}
\usepackage{epsf}
\usepackage{color}
\usepackage[utf8]{inputenc}

\begin{document}
\title[........1T-TaS$_2$]{Possible glass-like random singlet magnetic state in 1T-TaS$_2$}

\author{Sudip Pal, Kranti Kumar, Rohit Sharma, A Banerjee, and S B Roy}
 
\address{%
 UGC DAE Consortium for Scientific Research\\
 Khandwa Road, Indore 452001, India
}%

\author{Je-Geun Park}
\address{%
$^1$ Center for Correlated Electron Systems,\\
 Institute for Basic Science (IBS), Seoul 08826, Republic of Korea
}%
\address{%
 $^2$ Department of Physics and Astronomy, \\
Seoul National University, Seoul 08826, Republic of Korea
}%

\author{A K. Nigam}
\address{%
Tata Institute of Fundamental Research\\
 Mumbai 400005, India
}%

\author{Sang-Wook Cheong}
\address{%
Rutgers Center for Emergent Materials and Department of Physics and Astronomy\\
 Rutgers University, Piscataway, New Jersey 08854, USA
}%

\ead{sbroy@csr.res.in}
\vspace{10pt}
\begin{indented}
\item[]June 2019
\end{indented}

\begin{abstract}
Two-dimensional layered transition-metal-dichalcogenide compound 1T-TaS$_2$ shows the rare coexistence of charge density wave (CDW) and electron correlation driven Mott transition. In addition, atomic-cluster spins on the triangular lattice of the CDW state of 1T-TaS$_2$ give rise to the possibility of the exotic spin-singlet state  in which quantum fluctuations of spins are strong enough to prevent any long range magnetic ordering down to the temperature absolute zero ( 0 K). We present here the evidences of a glass-like random singlet magnetic state in 1T-TaS$_2$ at low temperatures through a study of temperature and time dependence of magnetization. Comparing the experimental results with a representative canonical spin-glass system Au(1.8$\%$Mn), we show that this glass-like state is distinctly different from the well established canonical spin-glass state.
\end{abstract}

\section{Introduction: }
The interplay of spin, lattice and charge degrees of freedom in 1T-TaS$_2$ gives rise to two interesting phenomena, which usually are not found together in the same system : the charge density wave (CDW) state and electron correlation driven Mott transition {\color{blue}\cite{fazekas1}}. This compound has a quasi-2D structure with each layer consisting of a triangular lattice of Ta atoms, which in turn is sandwiched by S atoms in an octahedral coordination. The CDW superstructure in 1T-TaS$_2$ (commonly known as commensurate CDW or CCDW state) has the basic units of so-called star-of-David clusters consisting of 13 Ta atoms and these clusters are ordered in a triangular lattice below 180 K. The CDW state, however, continues to exist even above 180 K in the form of a domain-like nearly commensurate CDW (or NC-CDW) state, which turns into  an incommensurate CDW (or IC-CDW) state above 340 K. The compound becomes a metal above 543 K. In the low-temperature CCDW state, 12 out of 13 Ta$^{4+}$ 5$\it d$-electrons star-of-David clusters form molecular orbitals, while leaving one 5$\it d$-electron with S = $\frac{1}{2}$ spin free. This leads to the formation of a very narrow band near the Fermi surface due to spin-orbit coupling {\color{blue}\cite{fazekas1,rossnagel}}, and a residual electron-electron interaction is enough to open a Mott gap {\color{blue}\cite{he}}.

In a separate development  in early 1970s,  P W Anderson proposed that a perfect triangular lattice like the orphan spin out of 13 Ta atoms in 1T-TaS$_2$ is a likely host of resonating valence bond state as the ground state of the triangular-lattice S =1$/$2 Heisenberg antiferromagnet instead of a more conventional Neel antiferromagnetic state {\color{blue}\cite{anderson}}. To this end there has been a significant interest in recent times  in the ground state properties of 1T-TaS$_2$ including the possible  existence of a quantum spin liquid (QSL), which is a state without spontaneously broken triangular-lattice symmetry and whose behaviour is dominated by emergent fractional excitations {\color{blue}\cite{he,lawlee}}. In a standard Mott insulator, the spins form local moments, which often forms an antiferromagnetic (AF) ordered state at lower temperature due to exchange coupling. However, there is no report so far of AF ordering in 1T-TaS$_2$, even there is no clear cut sign of the local moment formation. Magnetic susceptibility remains reasonably flat below approximately 200 K and down to 50 K, then rises monotonically down to the lowest measurable temperature {\color{blue}\cite{desalvo,ribak,kratoch}}. Law and Lee argued on the basis of analyzing these existing experimental results on 1T-TaS$_2$ that low temperature state may be considered either as a fully gaped Z2  spin liquid or a Dirac spin liquid {\color{blue}\cite{he,lawlee}}. We note here that there are conflicting reports on the existence of itinerant magnetic excitations {\color{blue}\cite{yu,mura}}.  Subsequent muon spin relaxation ($\mu$SR) and polarized neutron diffraction measurements indicated the presence of short-ranged magnetic order below 50 K without, however, any evidence of a long range magnetic order down to 70 mK  {\color{blue}\cite{kratoch}}. A  nuclear quadrupole resonance (NQR) experiment also revealed QSL-like behavior in 1T-TaS$_2$ from 200 K to 55 K, below which there were evidences of a novel quantum phase with amorphous tiling of frozen singlets emerging out of the QSL {\color{blue}\cite{klanjs}}.

In recent times it has also been reported that the low temperature state in 1T-TaS$_2$ is quite susceptible to external perturbations. Ultra fast resistance switching to a new CDW state with a lower resistance state can be induced by a 35 fs laser pulse {\color{blue}\cite{stojch}}, and also by a 40 ps electrical pulse-injection {\color{blue}\cite{vaski}}. This new state is termed as a hidden-CDW or HCDW state, which is quite robust in nature and can only be erased by heating above 80 K {\color{blue}\cite{svetin,wu}}. It is also reported recently that transient processes with rapid decay dynamics are triggered  upon weak pumping in time-domain terahertz measurements  in both CCDW and HCDW states   {\color{blue}\cite{zxwang}}. All these results have stimulated us to investigate carefully the stability of the low temperature state of 1T-TaS$_2$, and we report here the results of careful magnetization study of a single crystal and a polycrystal sample of 1T-TaS$_2$ subjected to various temperature and magnetic field cycling. Our results show that the low temperature magnetic state of 1T-TaS$_2$ is indeed metastable in nature, and this state is possibly an example of random singlet phase {\color{blue}\cite{fisher}}

\section{Sample Preparation : }
The single crystal sample of 1T-TaS$_2$ used in the present study is from the same batch of samples used earlier in various other studies including $\mu$SR and polarized neutron diffraction measurements {\color{blue}\cite{kratoch,yu,yu2}}, whereas the polycrystalline sample has been prepared freshly following the standard  solid state route {\color{blue}\cite{kratoch}}. This polycrystalline 1T-TaS$_2$ sample has been characterized with X-ray diffraction and EDAX studies, which ensured the single phase nature of the sample with the trigonal crystal structure having space group P$\bar3$m1. The spin-glass sample Au(1.8$\%$Mn) investigated here had been used in earlier studies {\color{blue}\cite{nigam}}, and readers are referred to that work for the details of sample preparation and characterization. It may be noted that some physical properties of 1T-TaS$_2$ depend on the thickness of the sample {\color{blue}\cite{yu2}}. However, the earlier reported temperature dependent resistivity study on the single crystal 1T-TaS$_2$ sample {\color{blue}\cite{yu}} used here, clearly indicates that the properties of this flake like single crystal are same as that of bulk 1T-TaS$_2$.  The magnetic measurements down to 2 K have been performed using a SQUID magnetometer (Quantum Design, USA) using a 4 cm scan length, as well as a vibrating sample magnetometer (Quantum Design, USA).

\begin{figure}[h]
\centering
\includegraphics[scale=0.3]{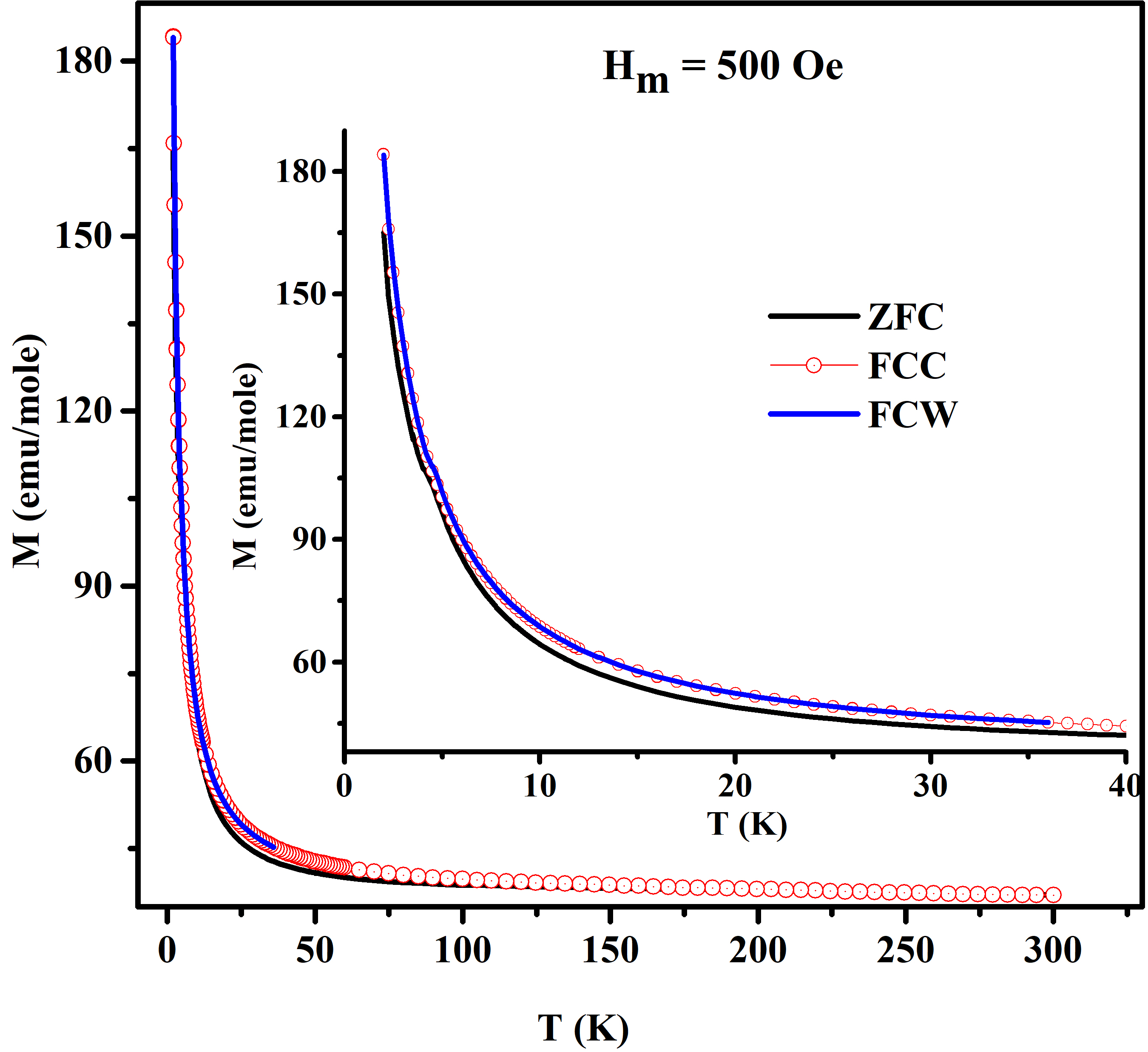}
\caption{ M vs T plots for 1T-TaS$_2$ obtained in ZFC, FCC and FCW mode (see text for details)  in the presence of an applied field of 500 Oe. The inset highlights the thermomagnetic irreversibility observed in the T regime below 50 K .}
\end{figure}

\section{\label{sec:level}Results and Discussions:\protect\\ }

Fig.1 shows the magnetization (M) versus temperature plot for the single crystal sample of 1T-TaS$_2$ obtained in 0.25 K steps in the temperature interval 2-12 K, in 1 K steps up to 60 K and then in 5 K steps up to room temperature, in zero field cooled (ZFC), field cooled cooling (FCC) and field cooled warming (FCW) mode in the presence of an applied magnetic field (H) of 500 Oe. In the ZFC mode the sample is cooled to the lowest measured temperature (here 2 K) before the applied H is switched on, and the measurement is made while warming up the sample. In the FCC mode the applied H is switched on at T = 300 K and the measurement is made while cooling to 2 K. After completion of measurements in the FCC mode, the data points are taken again in the presence of same applied H while warming up the sample. This is called FCW mode. The magnetization data obtained in the FCC and FCW mode merge at all temperatures except in a small temperature region around 200 K, which is not visible in the sacle of Fig.1. However, a distinct thermo-magnetic history effect between ZFC and FCC(FCW) mode i.e. M(T)$_{ZFC} \neq$ M(T)$_{FCC(W)}$ is observed below 50 K where the magnetization shows a rapid rise (please see the inset of Fig.1).  It may be noted that magnetization in this single crystal sample retained a small positive value right up to 300 K, which is possibly due to a small amount of 2H-polymorph present in the sample {\color{blue}\cite{stojch}}. The observed thermo-magnetic irreversibility below 50 K is present very much in the polycrystalline sample of 1T- TaS$_2$ we have studied (the results are not shown here for the sake of clarity and conciseness). This polycrystalline sample shows the expected diamagnetic behaviour in the high temperature regime, and thus rules out any impurity related origin of the observed thermo-magnetic irreversibility. We have also ruled out the possibility of any extraneous origin of the observed behaviour by repeating some of the experiments using another magnetometer namely a vibrating sample magnetometer (VSM), and in the rest of the paper we will only present results obtained on the single crystal sample of 1T-TaS$_2$ using the SQUID magnetometer. In an expanded scale in the Fig.1, the higher temperature NC-CDW to CCDW transition in 1T-TaS$_2$ is also visible in the form of a subtle change in slope in the magnetization curve accompanied by a thermal hysteresis between M$_{FCW}$(M$_{ZFC}$) and M$_{FCC}$, but that is not shown here for the sake of clarity and conciseness. It may be noted here that  no thermal hysteresis is observed in the temperature dependence of magnetization obtained in the FCW and FCC mode at any other temperature regime.

\begin{figure}[h]
\centering
\includegraphics[scale=0.3]{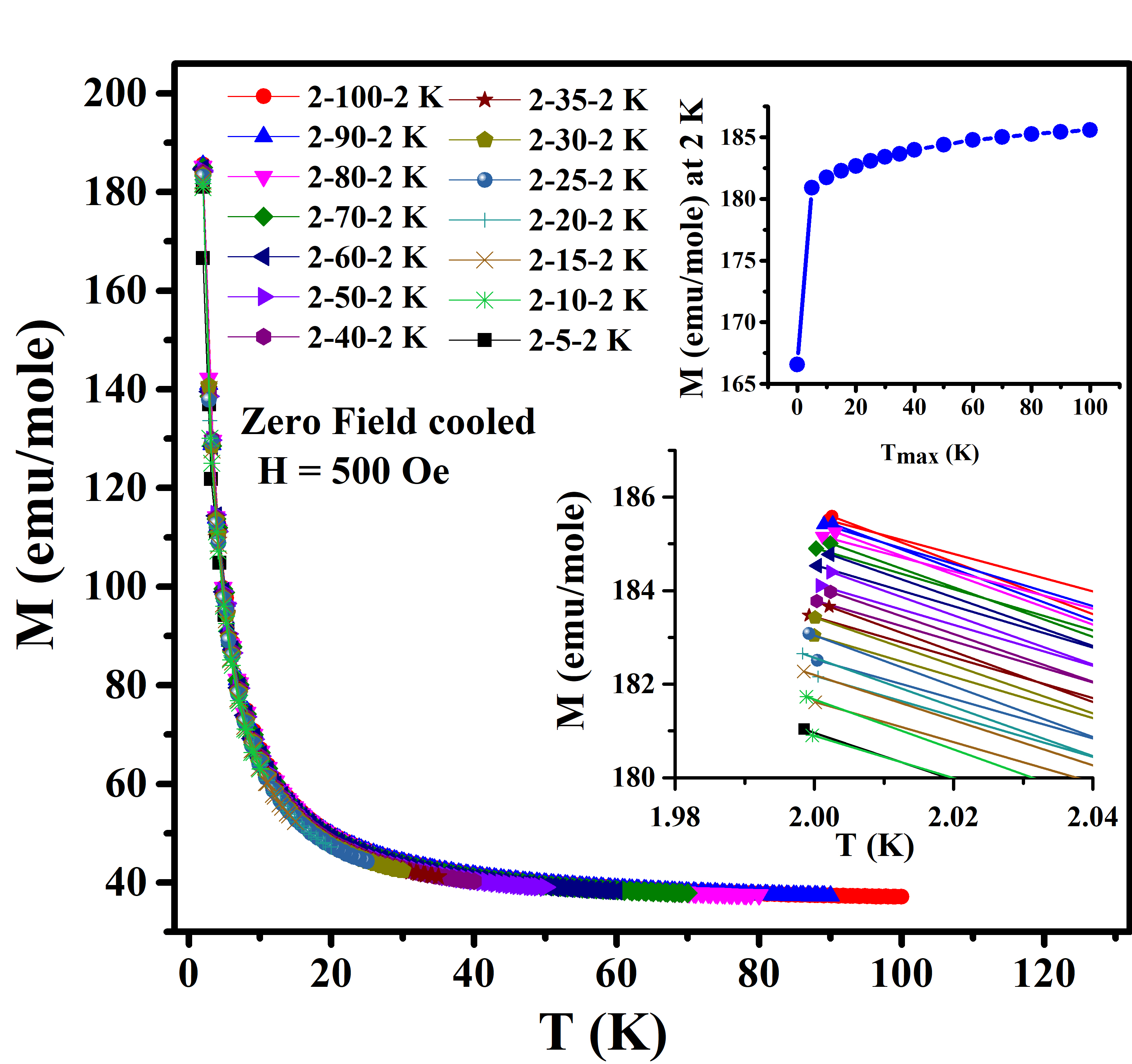}
\caption{ M-T plots showing the effect of thermal cycling on the ZFC state of 1T-TaS$_2$ to highlight the metastable behaviour of this state. The lower inset shows how the magnetization value progressively increases at 2 K  after each thermal cycling.  Here the magnetization is measured down to 2 K after repeated excursions to progressively higher temperatures (see the text and legends in the figure for details). The upper inset plots the magnitude of magnetization at 2 K after temperature cycling from various  temperatures T$_{max} \geq$ 5 K.}
\end{figure}

Fig.2 presents the results on the effect of thermal cycling in the ZFC state of 1T-TaS$_2$, highlighting the metastable nature of this low temperature ZFC state. In this experiment, the measurement was done on the 1T-TaS$_2$ sample in the ZFC mode by increasing the temperature and then returning back to 2 K repeatedly with 5 K temperature  step in the temperature range 2 K $\leq$ T $\leq$ 40 K and then in 10 K step up to 100 K. The magnetization at 2 K value is found to be increasing from the starting value of the ZFC mode with each thermal cycling steps (see insets of Fig.2). The value of magnetization actually increased quite abruptly in the first 5 K step, which is followed by a rather monotonic increase to a saturation value above 50 K. We had continued with our experiment with bigger steps of 10 K up to a temperature of 100 K; no effect of thermal cycling is observed above 50 K, and the magnetization obtained in the ZFC and FC mode merges together. 

\begin{figure}[h]
\centering
\includegraphics[scale=0.3]{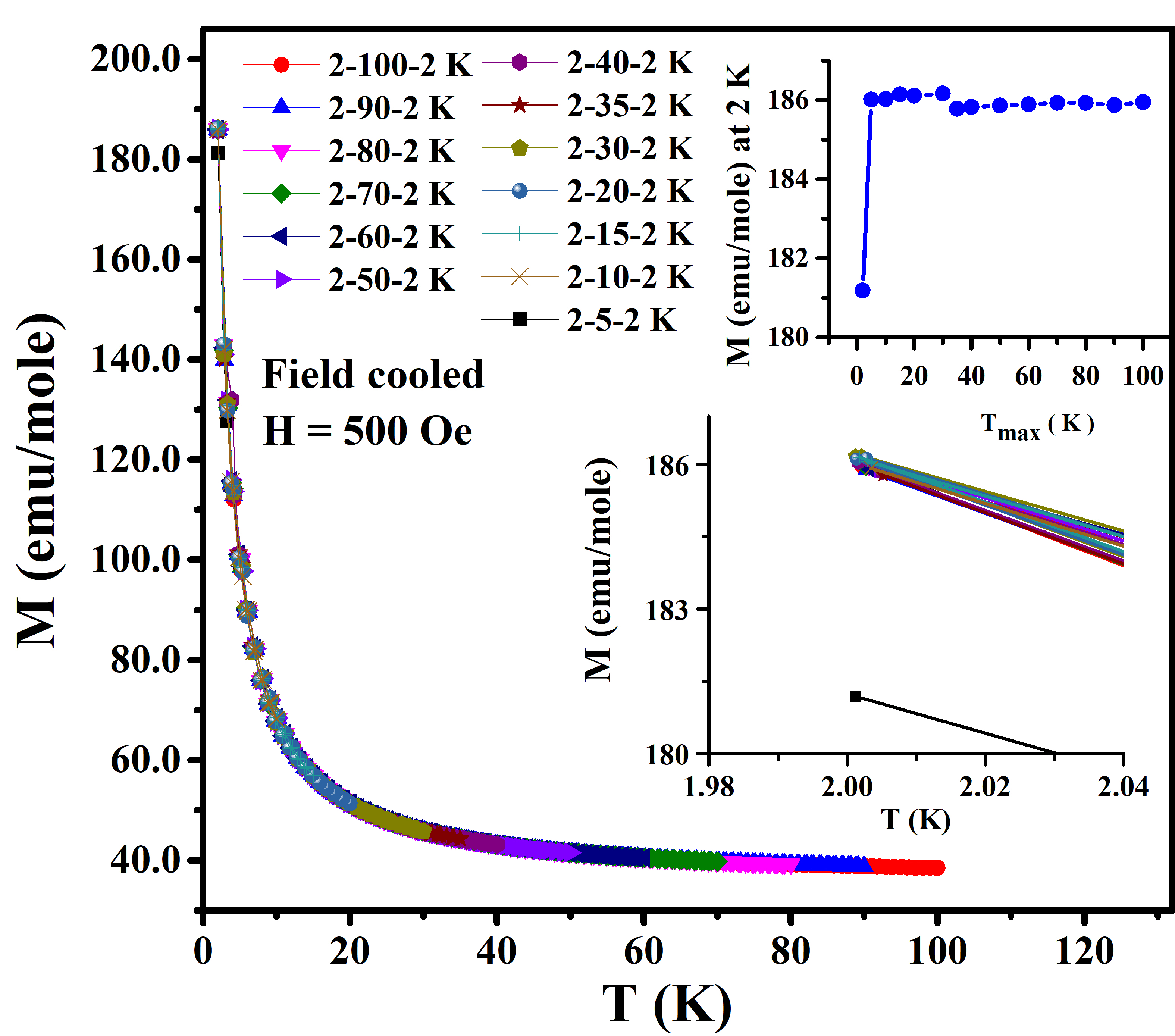}
\caption{ M-T plots showing the effect of thermal cycling on the FC state of 1T-TaS$_2$ to highlight the metastable behaviour of this state. The lower inset shows the response of magnetization at 2 K after thermal cycling, where the magnetization is measured down to 2 K after repeated excursions to progressively higher temperatures (see the text and legends in the figure for details). The magnetization increases after the first thermal cycling, and thereafter it remains reasonably constant. The upper inset plots the magnitude of magnetization at 2 K after temperature cycling from various  temperatures T$_{max} \geq$ 5 K.}
\end{figure}

Fig.3 presents the results of thermal cycling in the FC state of 1T-TaS$_2$, highlighting a subtle metastable nature of the low temperature FC state too. The experimental protocol followed is the same as in the case of ZFC mode, except that the initial cooling from room temperature to 2 K took place in the presence of an applied field of 500 Oe. The extent of metastability, however, is distinctly different from that observed in the ZFC state. Like in the case of ZFC state, the value of magnetization increased in the first 5 K temperature cycling step,  but the magnitude of this increase is significantly smaller than that observed in the ZFC state (see lower insets of Fig.2 and 3). It may be noted here that the starting value of magnetization  at 2 K  itself is higher in the FC state  than that in the ZFC state. There is no further significant change in magnetization in the FC-state at 2 K on subsequent temperature cycling, and the magnetization retains a saturation value (see upper inset of Fig.3).  It may be recalled here that while similar thermo-magnetic irreversibility i.e  M(T)$_{ZFC} \neq$ M(T)$_{FC}$ is a hallmark of spin-glass like state, but there it is generally accepted that the FC-state is an equilibrium state. Hence, the observed subtle metastable behaviour in the FC-state of 1T-TaS$_2$ clearly distinguishes this compound from a canonical spin-glass.

\begin{figure}[h]
\centering
\includegraphics[scale=0.4]{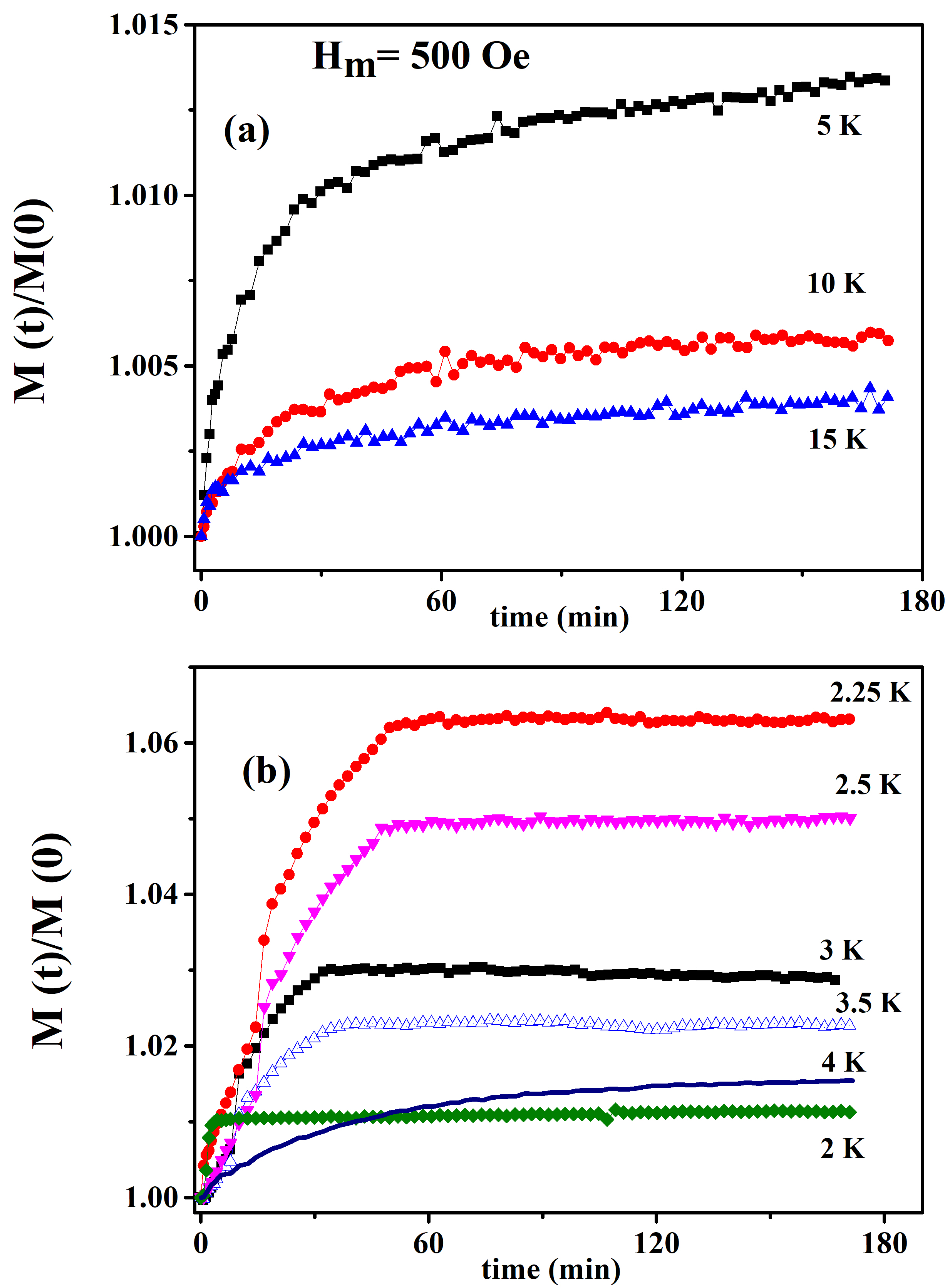}
\caption{Magnetization versus time plots in the ZFC state of  1T-TaS$_2$ obtained in an applied magnetic field of 500 Oe. It may be noted that the nature of magnetization relaxation changes drastically in the temperature region below 4 K (see Fig. 4(b)) from that of the higher temperature ( T $>$ 4 K) regime (see Fig. 4(a)).}
\end{figure}

This metastable response in the low temperature state of 1T-TaS$_2$ is further investigated through magnetic relaxation measurements. In this experiment magnetization is studied in the ZFC mode as a function of time after stabilizing the sample in various temperatures below 50 K (see Fig.4). This revealed a rather unusual behaviour of magnetization relaxation with decreasing temperature. First of all, the observed relaxation in magnetization (M) indicates ageing in the system with a broad distribution of relaxation rates.  The variation of M as a function time can be fitted reasonably well with a stretched exponential function having three free parameters. However, these relaxation results can be represented even better with a logarithmic time growth of the magnetization (see Fig.5)  as:

\begin{equation}
M(t) = M(0)[1 + Dln(t/\tau)]
\end{equation}

In this eqn.1, the term D is known as rate constant {\color{blue}\cite{alok}}.This kind of logarithmic function has been successfully used to study non-equilibrium phenomena in various kinds of systems including structural-glass, spin-glass and shape memory alloys {\color{blue}\cite{alok,amir}}.  This logarithmic growth of M in the ZFC state of 1T-TaS$_2$ is observed down to 4 K. In Fig.5 we compare this with the results of a similar study on the ZFC state of a canonical spin-glass namely Au(1.8$\%$Mn) with a spin-glass transition temperature around T$_{sg}$ = 7 K. The rate constant D of the two systems are compared in the inset of Fig.5.  The rate constant of the spin-glass system shows a broad maximum as a function of temperature. In 1T-TaS$_2$ rate constant increases monotonically until 4 K. In the light of relatively more metastable nature of the ZFC state of 1T-TaS$_2$, the relaxation experiments are  mostly performed in the ZFC state.

 A drastic change in the behaviour of magnetization relaxation takes place in 1T-TaS$_2$ below 4 K. The magnetization shows a rapid increase with time before saturating to a constant value (see Fig. 4(b)). The time dependence of magnetization can no longer be fitted with eqn.1, and the rate constant D loses its meaning in the temperature region below 4 K. This rather anomalous time dependence of magnetization is observed also in the FC state. Very similar behavior of magnetization relaxation has been reported for  Dy$_2$Ti$_2$O$_7$  below 600 mK, which was attributed to monopole dynamics {\color{blue}\cite{paulsen}}. Earlier, a rise in spin relaxation below 4 K in Dy$_2$Ti$_2$O$_7$  has also been attributed to the emergence of a collective degree of freedom for which thermal relaxation processes is important as the spins become strongly correlated {\color{blue}\cite{snyder}}.  The observed similar behaviour in 1T-TaS$_2$ may thus indicate the emergence of a collective degree of freedom with strongly correlated spins. 

\begin{figure}[h]
\centering
\includegraphics[scale=0.4]{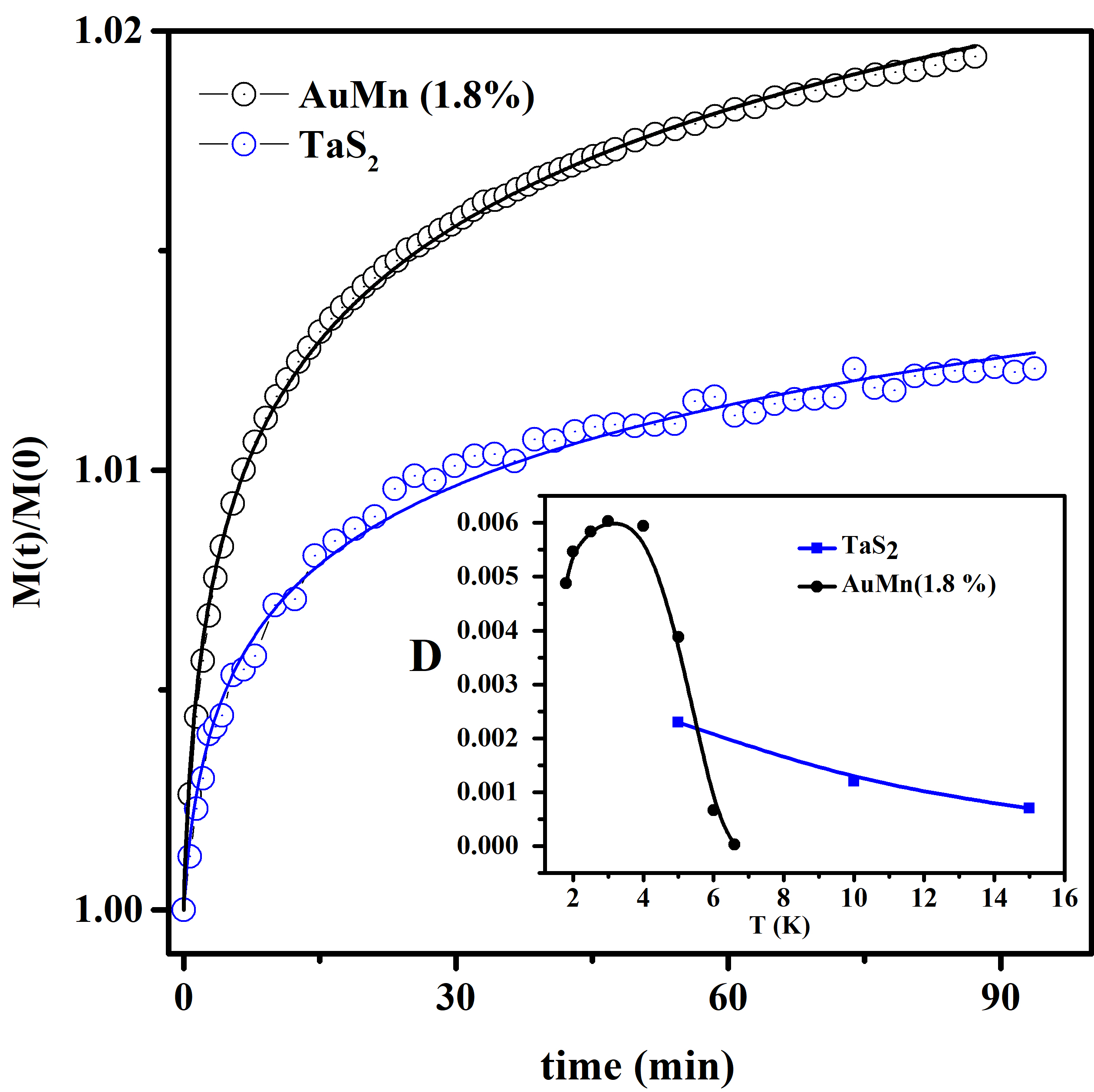}
\caption{ The logarithmic growth of magnetization in the ZFC state of spin-glass Au(1.8$\%$Mn) and 1T-TaS$_2$ measured at T = 5 K. The inset shows variation of the rate constant D for the two samples as a function of temperature. The solid lines in the main frame figure present the theoretical fit to the equation 1, whereas the lines in the inset represent guideline to the data points. It may be noted that below 4 K the time dependence of magnetization in 1T-TaS$_2$ can no longer be fitted with eqn.1, and the rate constant D loses its meaning.}
\end{figure}

The observed anomalous magnetization relaxation below 4 K in 1T-TaS$_2$ can, however,  have origin in a possible experimental artifact. It may be noted that the magnetization in 1T-TaS$_2$ increases very rapidly with temperature in the temperature regime below 10 K.  The time relaxation measurements  reported here have been performed by cooling the sample in zero field from room temperature with a cooling rate of 1.5 K$/$min and then stabilizing the temperature at the temperature of measurement. Although the display in the SQUID magnetometer indicated a stable temperature, in reality a small temperature drift  may continue to exist at the sample site, which might lead to a variation of magnetization as a function of time. To consider this possibility we have checked the results at some temperatures above 2 K by cooling the sample first to 2 K in the ZFC mode, and then by increasing the temperature slowly to the targeted temperature of measurements before switching on the measuring field. However, these measurements could not rule out unequivocally the effect of a possible but yet unknown/undetected temperature drift. This 'undetected temperature drift' assumption, however, has certain problem. The observed increase in magnetization in logarithmic time scale in the temperature regime above 4 K in 1T-TaS$_2$ then would imply that the temperature in the SQUID magnetometer always keeps on drifting uni-directionally downwards in logarithmic time scale. This is quite unlikely with a standard PID temperature controller employed in such commercial magnetometers.    

We have performed another set measurement by cooling the sample from room temperature to 60 K first with the cooling rate of 1.5 K $/$ min and then waiting there for 6 hrs before cooling down to various temperatures of measurements. In this mode of measurement the observed anomalous time dependence of magnetization at temperature below 4 K totally disappeared. However, there is a very recent report on the cooling rate dependence of the low temperature state of 1T-TaS$_2$, which indicated that the thermodynamic ground state can be reached only with very slow cooling {\color{blue}\cite{lee}}. Hence the waiting time of 6 hrs at 60 K in our measurement, possibly caused the sample to deviate from the state that was achieved by cooling the sample directly to the  lower temperature of measurements.

As mentioned in the beginning of the paper, the idea of quantum spin-liquid involving a concept of ‘resonating valence bond’ was due to P W Anderson  {\color{blue}\cite{anderson}}. Two neighbouring spins interacting antiferromagnetically can pair into a singlet state, and form a valence bond or dimer. The ground state can be represented by the product of the valence bonds when all the spins in a system form valence bonds.  This is a valence bond solid (VBS) (see Fig. 6(a)). A VBS state, however, is not a quantum spin-liquid, since it can break lattice symmetries and it lacks long-range entanglement  {\color{blue}\cite{balent}}. In order to reach the quantum spin-liquid state, the valence bonds must be allowed to undergo quantum mechanical fluctuations. Here Anderson invoked the idea of a superposition of VBS states, which was earlier named by Linus Pauling as a resonating valence bond (RVB)  {\color{blue}\cite{pauling}} . Anderson  {\color{blue}\cite{anderson,anderson87}} proposed that in the triangular two dimensional spin-1/2 antiferromagnet the ground state is analogous to the precise singlet in the Bethe solution of the linear antiferromagnetic chain. Instead of forming a fixed array of spin singlets, in such cases strong quantum fluctuations lead to a superposition of singlet configurations (see Fig. 6(b)). 

\begin{figure}[h]
\centering
\includegraphics[scale=0.5]{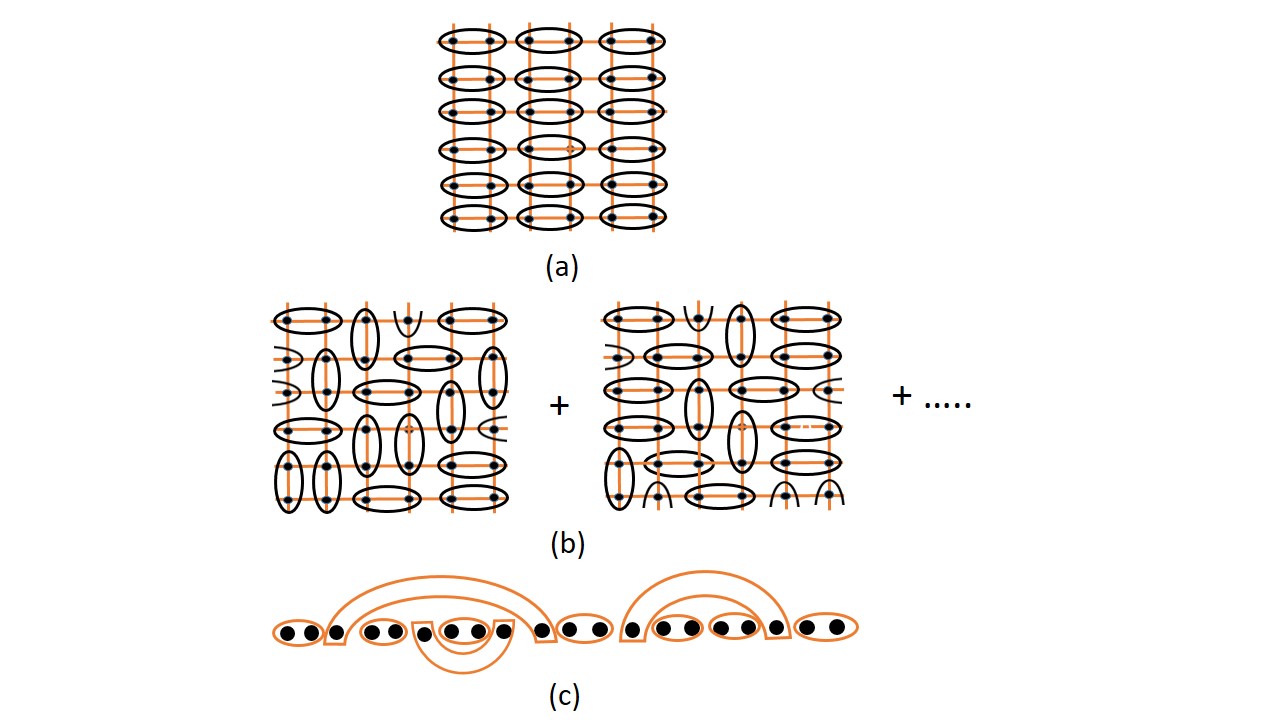}
\caption{ Schemtaic representation of: (a) valence bond solid (VBS) state; (b) resonating valence bond (RVB) state; (c) random singlet state with short and long singlet bonds.}
\end{figure}

In this light it is possible to rationalize the observed results in 1T-TaS$_2$ at low temperatures within the framework of a random-singlet phase {\color{blue}\cite{fisher}}.  In this phase the system at low energies can have pairs of spins which are coupled together into singlets over arbitrarily long distances (see Fig. 6(c)). The long singlet bonds (represented by large distorted ellipses in fig. 6(c)) are expected to be much weaker than the short singlet bonds (represented by small ellipses in Fig. 6(c)),  and the singlet bonds cannot cross. In the event of the distribution of bare exchange couplings $\it J$ being narrow, the physics at energies of order the initial $\it J$'s would cause the spin-1$/$2 objects forming the low-energy singlets to spread out over a number of lattice sites {\color{blue}\cite{fisher}}. The observed increase of large relaxation rate is indicative of forming of long singlet bond. The system still remains in a glass-like state involving a few numbers of long singlet bond, which may be distinguished from higher temperature glass-like state involving more numbers of nearest neighbour or short singlet bonds. The experimental features of the glass-like random singlet states in 1T-TaS$_2$ are clearly different from the canonical spin-glass systems, as we have exemplified by comparing the relaxation results obtained in canonical spin-glass system Au(1.8$\%$Mn). Furthermore, unlike in spin-glasses the FC state of 1T-TaS$_2$ seems to have subtle metastability.  The present experimental study clearly indicates that both the low temperature ZFC and FC states in 1T-TaS$_2$ are metastable in nature. This in turn points out that the low temperature state of 1T-TaS$_2$ is not really in thermal equilibrium,  or the system may need some finite amount of time to reach the equilibrium state. The question still remains here whether this low temperatures state is really the ground state of 1T-TaS$_2$ or one of the many possible quasi-equilibrium states, which are close in energy. In fact, there exist some experimental results, which indicate such a possibility in 1T-TaS$_2$ {\color{blue}\cite{yoshida}}. There are also some theoretical efforts to investigate the effect of disorder and randomness in VBS systems {\color{blue}\cite{watan,kawa,tok,uem,kimchi}}. It has been argued that the quenched randomness leads to the possibilities of new kinds of quantum ground states with interesting entanglement structures. One of the possibilities is that the weak disorder transforms a paramagnetic valence-bond solid into a state with 'spinful excitations', starting with the nucleation of spin-1$/$2 vortex defects {\color{blue}\cite{kimchi}}. In the stronger disorder regime the spin-1/2 defects may lead to a 'glassy' covering of short-range valence bonds rather than an ordered one {\color{blue}\cite{kimchi}}. However, in the present case of 1T-TaS$_2$ the absence of any kink or change in slope in the temperature dependence of low field  magnetization along with the  presence of a subtle-metastability even in the FC state, suggest that the observed glass like features may not be entirely due to the frustrated spin-spin interactions as in the canonical spin-glasses. The structural aspects along with the possibility of spin-lattice interaction seem to be playing an important role here. The source of randomness in 1T-TaS$_2$, however, is not quite clear, and can be due to stacking faults in TaS$_2$ planes or even slight off-stoichiometry of the sample {\color{blue}\cite{klanjs}}.

\section{\label{sec:level}Acknowledgement:\protect\\ }
We thank R. Rawat for providing facility to synthesize polycrystalline sample of 1T-TaS$_2$ used in this study, and Pavel Volkov and Marie Kratochvilova for useful discussions.  The authors at UGC-DAE CSR, Indore thank Director and Centre Director for support and encouragement. The work in Korea was supported by the Institute for Basic Science (IBS) in Korea (IBS-R009-G1). The work at Rutgers University was supported by the NSF under Grant No. NSF-DMREF-DMR-1629059.

\end{document}